\documentstyle{pasj06}
\begin{document}

\SetRunningHead{A.Imada et al.}{}

\title{The 2006 November outburst of EG Aquarii: the SU UMa nature revealed}

\author{
Akira \textsc{Imada}\altaffilmark{1},
Rod \textsc{Stubbings}\altaffilmark{2},
Taichi \textsc{Kato}\altaffilmark{1},
Makoto \textsc{Uemura}\altaffilmark{3},
Thomas \textsc{Krajci}\altaffilmark{4}, \\
Ken'ichi \textsc{Torii}\altaffilmark{5}, 
Kei \textsc{Sugiyasu}\altaffilmark{1},
Kaori \textsc{Kubota}\altaffilmark{1},
Yuuki \textsc{Moritani}\altaffilmark{1}
Ryoko \textsc{Ishioka}\altaffilmark{6}, \\
Gianluca \textsc{Masi}\altaffilmark{7}, 
Seiichiro \textsc{Kiyota}\altaffilmark{8}, 
L.A.G. \textsc{Monard}\altaffilmark{9},
Hiroyuki \textsc{Maehara}\altaffilmark{10},\\
Kazuhiro \textsc{Nakajima}\altaffilmark{11}, 
Akira \textsc{Arai}\altaffilmark{12},
Takashi \textsc{Ohsugi}\altaffilmark{3}, 
Takuya \textsc{Yamashita}\altaffilmark{3},\\
Koji S. \textsc{Kawabata}\altaffilmark{3}, 
Osamu \textsc{Nagae}\altaffilmark{12}, 
Shingo \textsc{Chiyonobu}\altaffilmark{12},
Yasushi \textsc{Fukazawa}\altaffilmark{12}, \\ 
Tsunefumi \textsc{Mizuno}\altaffilmark{12},
Hideaki \textsc{Katagiri}\altaffilmark{12},
Hiromitsu \textsc{Takahashi}\altaffilmark{12}, 
Atsushi \textsc{Ueda}\altaffilmark{12}, \\
Takehiro \textsc{Hayashi}\altaffilmark{13}, 
Kiichi \textsc{Okita}\altaffilmark{14},
Michitoshi \textsc{Yoshida}\altaffilmark{14},
Kenshi \textsc{Yanagisawa}\altaffilmark{14}, \\ 
Shuji \textsc{Sato}\altaffilmark{15}, 
Masaru \textsc{Kino}\altaffilmark{15}, 
Masahiro \textsc{Kitagawa}\altaffilmark{15}, 
Kozo \textsc{Sadakane}\altaffilmark{16} \\
and Daisaku \textsc{Nogami}\altaffilmark{17}
}

\affil{$^1$Department of Astronomy,Faculty of Science, Kyoto University,
       Sakyo-ku, Kyoto 606-8502}
\affil{$^2$Tetoora Observatory, Tetoora Road, Victoria, Australia}
\affil{$^3$Astrophysical Science Center, Hiroshima University, Kagamiyama
1-3-1, \\ Higashi-Hiroshima 739-8526}
\affil{$^4$PO Box 1351 Cloudcroft, New Mexico 83117, USA}
\affil{$^5$Department of Earth and Space Science, Graduate School of
Science, \\ Osaka University, 1-1 Machikaneyama-cho, Toyonaka, Osaka
       560-0043}
\affil{$^6$Subaru Telescope, National Astronomical Observatory of Japan, 650
North A'ohoku Place,\\ Hilo, HI 96720, USA}
\affil{$^7$The Virtual Telescope Project, Via Madonna del Loco 47, 03023
       Ceccano (FR), Italy}
\affil{$^8$VSOLJ, Center for Balcony Astrophysics, 1-401-810 Azuma, Tsukuba,
Ibaraki 305-0031}
\affil{$^9$Bronberg Observatory, CBA Pretoria, PO Box 11426, Tiegerpoort
       0056, South Africa}
\affil{$^{10}$Kwasan and Hida Observatories, Kyoto University,
       Yamashina-ku, Kyoto 607-8471}
\affil{$^{11}$VSOLJ, 124 Isatotyo Teradani, Kumano, Mie 519-4673}
\altaffiltext{12}{Department of Physical Science, Hiroshima University,
Kagamiyama 1-3-1, \\ Higashi-Hiroshima 739-8526}
\altaffiltext{13}{Department of Education, Hiroshima University,
Kagamiyama 1-1-1, \\ Higashi-Hiroshima 739-8524}
\altaffiltext{14}{Okayama Astrophysical Observatory, National
Astronomical Observatory of Japan, \\ Kamogata Okayama 719-0232}
\altaffiltext{15}{Department of Physics, Nagoya University, Furo-cho,
Chikusa-ku, Nagoya 464-8602}
\altaffiltext{16}{Astronomical Institute, Osaka Kyoiku University,
Asahigaoka, Kashiwara, Osaka 582-8582}
\affil{$^{17}$Hida Observatory, Kyoto University, Kamitakara, Gifu
       506-1314}
\email{a\_imada@kusastro.kyoto-u.ac.jp}


\maketitle

\begin{abstract}

We report time-resolved CCD photometry of the cataclysmic variable EG
 Aquarii during the 2006 November outburst During the outburst, superhumps
 were unambiguously detected with a mean period of 0.078828(6) days,
 firstly classifying the object as an SU UMa-type dwarf nova. It also
 turned out that the outburst contained a precursor. At the end of the
 precursor, immature profiles of humps were observed. By a phase
 analysis of these humps, we interpreted the features as superhumps. This
 is the second example that the superhumps were shown during a
 precursor. Near the maximum stage of the outburst, we discovered an
 abrupt shift of the superhump period by ${\sim}$ 0.002 days. After the
 supermaximum, the superhump period decreased at the rate of
 $\dot{P}/P$=$-$8.2${\times}$10$^{-5}$, which is typical for SU UMa-type
 dwarf novae. Although the outburst light curve was characteristic of SU
 UMa-type dwarf novae, long-term monitoring of the variable shows no
 outbursts over the past decade. We note on the basic properties of long
 period and inactive SU UMa-type dwarf novae.

\end{abstract}

\section{Introduction}

Dwarf novae are a subclass of cataclysmic variables (CVs) that consist
of a white dwarf primary and a late-type secondary. The secondary fills
its Roche lobe and transfers gases into the primary through the inner
Lagragian point (L1). The transferred matter forms an accretion disk
around the white dwarf. The accretion disk causes instabilities, which
are observed as an outburst (for a review, see \cite{war95book};
\cite{osa96review}; \cite{hel01book}; \cite{smi07review}).

SU UMa-type stars are a subclass of dwarf novae. The majority have
orbital periods below 2 hours. This
subclass exhibits two types of outbursts. One is a normal outburst,
lasting a few days. The other is a superoutburst lasting about 2
weeks. The most characteristic feature during the superoutburst is that
the light curve always shows modulations termed superhump. The observed
period of superhumps is a few percent longer than the
orbital period of the system. This is well explained by the precessing
eccentric disk, which is deformed at the 3:1 resonance radius
(\cite{whi88tidal}; \cite{osa89suuma}).

EG Aquarii (hereafter EG Aqr) was introduced as a candidate for dwarf
novae by \citet{1959PASP...71..469L} and later by
\citet{vog82atlas}. The variability of EG Aqr was reported in
\citet{har60egaqr}, in which the variable
was at 14.0 mag on 1958 November 5. There are no other positive and
negative observations except this around this outburst, so that we cannot
lead to the conclusion of the nature of the
outburst. \citet{szk92CVspec} performed optical spectroscopy of the
variable. The result was, however, that the spectrum showed no emission
in Balmer series. On the contrary, it resembles that of a K-type
star. There is a possibility that the star was simply misidentified by
\citet{szk92CVspec}. Our astrometric estimation of the variable from the
outbursting images yielded 
RA $23^{\rm h} 25^{\rm m} 19^{\rm s}.09$ and 
Dec $-08^{\circ} 18' 18''.5$, respectively. This means that the variable
is identical with USNO B1.0 0816-0716959 
(RA $23^{\rm h} 25^{\rm m} 19^{\rm s}.17$, Dec $-08^{\circ} 18' 18''.9$, 
$B1$=18.760 $R1$=18.260).\footnote{A. Henden gives $V$=18.651 and
$B-V$=-0.062.}. Concerning these results, the infrared counterpart may
be 2MASS J23251917-0818190 ($J$=17.432, $H$=16.309,
$K$=15.681.\footnote{\cite{spr96CVabsmag} gave $J$= 17.10 and $K$=16.26,
respectively. see also \citet{hoa02CV2MASS}).} As for the X-ray range,
the variable is below the detection limit of the ROSAT faint source
catalog \citep{1999A&A...349..389V}.

On 2006 November 8, one of the authors (RS) discovered the eruption
of EG Aqr at a visual magnitude of 12.5 ([vsnet-alert 7217]). This is
the first recorded outburst since the report by \citet{har60egaqr}. Here
we report on time-resolved CCD photometry of EG Aqr during the 2006
November superoutburst.

\section{Observations}

\begin{longtable}{cccccc}
\caption{Journal of observations.}
\hline\hline
2006 Date & HJD-start$^*$ & HJD-end$^*$ & Exp(s)$^{\dagger}$
& N$^{\ddagger}$ & ID$^{\S}$ \\
\hline
\endhead
\hline
\endfoot
 November 8 & 48.2507 & 48.2719 & 30 & 57 & BM \\
 & 48.3433 & 48.4689 & 60 & 161 & GM \\
 November 9 & 48.5502 & 48.7896 & 30 & 559 & KTC\\
 & 48.8877 & 48.9843 & 30 & 140 & Njh \\
 & 48.9031 & 49.0906 & 30 & 285 & Kyoto \\
 & 48.9345 & 49.0642 & 30 & 341 & Kis \\
 & 49.0311 & 49.1369 & 30 & 400 & Mhh \\
 & 49.2279 & 49.3960 & 60 & 212 & GM \\
 & 49.3272 & 49.4042 & 30 & 219 & BM \\
 November 10 & 49.5484 & 49.7925 & 30 & 570 & KTC\\
 & 50.2290 & 50.4390 & 60 & 233 & GM \\
 November 11 & 50.5497 & 50.8028 & 30 & 591 & KTC\\
 & 50.9310 & 50.9632 & 30 & 39 & Njh \\
 November 12 & 51.8683 & 52.0512 & 30 & 390 & Kyoto \\
 & 51.8824 & 52.0413 & 30 & 280 & Njh \\ 
 & 51.8851 & 51.9495 & 30 & 142 & Kis \\
 & 51.9244 & 52.1404 & 30 & 749 & Mhh \\
 November 13 & 52.5921 & 52.8031 & 30 & 493 & KTC\\
 & 52.9314 & 52.9988 & 30 & 109 & Njh \\
 & 53.0267 & 53.0671 & 30 &  71 & Kyoto \\
 Novemver 14 & 53.5503 & 53.8013 & 30 & 586 & KTC\\
 & 53.9061 & 54.0310 & 30 & 410 & Mhh \\
 November 15 & 54.8652 & 55.0377 & 30 & 227 & Kyoto \\
 & 54.9020 & 55.0546 & 30 & 281 & Njh \\
 & 55.2578 & 55.4017 & 60 & 177 & GM \\
 November 16 & 55.5533 & 55.8009 & 30 & 578 & KTC\\
 & 55.8599 & 56.0782 & 30 & 338 & Kyoto \\
 & 55.8835 & 56.0688 & 30 & 312 & Njh \\
November 17 & 56.6015 & 56.7992 & 30 & 462 & KTC \\
 & 56.8768 & 57.1001 & 30 & 350 & Kyoto \\
 & 56.8834 & 56.9822 & 30 & 180 & Njh \\
November 18 & 57.5465 & 57.7572 & 30 & 351 & KTC \\
 November 20 & 59.8828 & 60.0813 & 30 & 265 & Kyoto \\
November 21 & 61.0241 & 61.1100 & 63 & 81 & Kanata \\
November 24 & 63.9465 & 64.0325 & 10 & 606 & Kanata \\
December 5 & 74.9825 & 75.0276 & 63 & 48 & Kanata \\
December 6 & 76.0116 & 76.0383 & 63 & 26 & Kanata \\
\hline
\multicolumn{6}{l}{$^*$HJD-2454000. $^{\dagger}$Exposure
 time. $^{\ddagger}$Number of exposure. $^{\S}$Observer's ID.} \\
\multicolumn{6}{l}{BM: L.A.G. Monard, South Africa. GM: G. Masi, Italy.} \\
\multicolumn{6}{l}{KTC: T. Krajci, USA. Njh: K. Nakajima, Japan.} \\
\multicolumn{6}{l}{Kyoto: A. Imada et al., Japan. Kis: S. Kiyota, Japan.} \\
\multicolumn{6}{l}{Mhh: H. Maehara, Japan. Kanata: M. Uemura et al., Japan.} \\
\label{logofobs}
\end{longtable}

Time resolved CCD photometry was performed on 15 nights using 25-150cm
telescopes between 2006 November 8 and 2006 December 6 at 8 sites, and a
log of observations is summarized in table 1. No filter was used
except for Higashi-Hiroshima site and the obtained data were close to
those of
$R_{\rm c}-$system. At Higashi-Hiroshima site, the data were obtained
with the
Triple Range Imager and Spectrograph (TRISPEC,
\cite{2005PASP..117..870W}) using a $V$ filter on 2006 November 21, December
5, and December 6. The exposure time was 10-63 seconds, with a few second
readout time of the CCDs. The total datapoint amounts to 10506, which is
sufficient to investigate the variability of the system.

After debiasing and flat fielding, the images were processed by aperture
photometry packages. Since there existed an effect of the atmospheric
extinction due to the color difference between the variable and the
comparison star, we made a correction for the data. After correcting the
obtained magnitudes between sites, the magnitude was adjusted to that of the
Kyoto site, where we used a Java-based aperture photometry package
developed by one of the authors (TK). The differential magnitude of the
variables were determined using No. 12 of the Henden Catalog ($V$=11.910,
$B-V$=0.632), whose constancy was checked using nearby stars in the same
images.\footnote{ftp://ftp.aavso.org/public/calib/} The accuracy of the
calibration was dependent on the condition of the skies. Although some
data were contaminated by such as clowds or light pollutions, the
expected error was achieved as small as 0.03 mag for good data, which is
sufficient for the purpose of our observations. Heliocentric correction
was made for each data set before the following analyses.

\section{Results}

\begin{figure*}
\begin{center}
\resizebox{160mm}{!}{\includegraphics{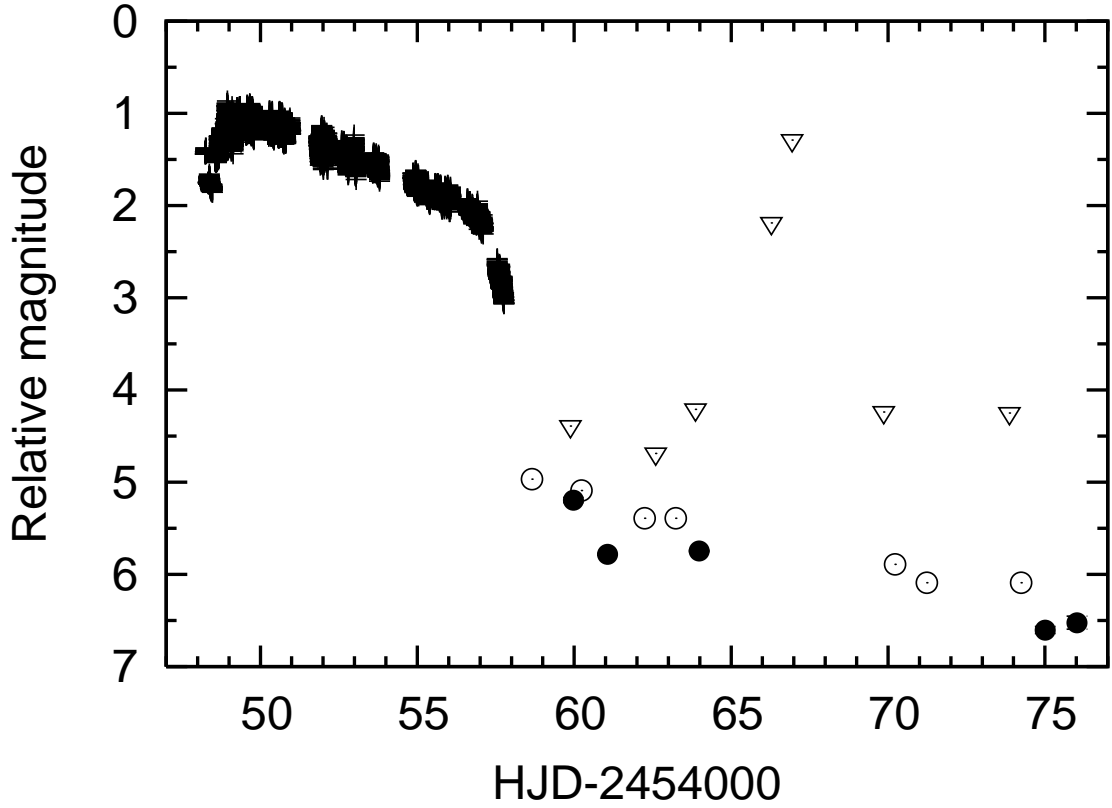}}
\end{center}
\caption{Light curve of EG Aqr during the 2006 November
 superoutburst. The abscissa and the ordinate denote HJD and relative
 magnitude, respectively. The comparison star is at a magnitude of 11.91
 in $V$ band. The filled and open circles indicate an averaged light curves
 obtained by our run and the AAVSO database, respectively. The bottom
 triangles denotes negative observations reported to the AAVSO and
 VSNET. A precursor was detected on HJD 2454048 (see also figure 2). The
 maximum magnitude was reached around HJD 2454049.7, after that the
 variable declined at the rate of 0.14(1) mag d$^{-1}$. No evidence for
 a rebrightening was shown during our whole run.}
\label{lc}
\end{figure*}

\begin{figure*}
\begin{center}
\resizebox{160mm}{!}{\includegraphics{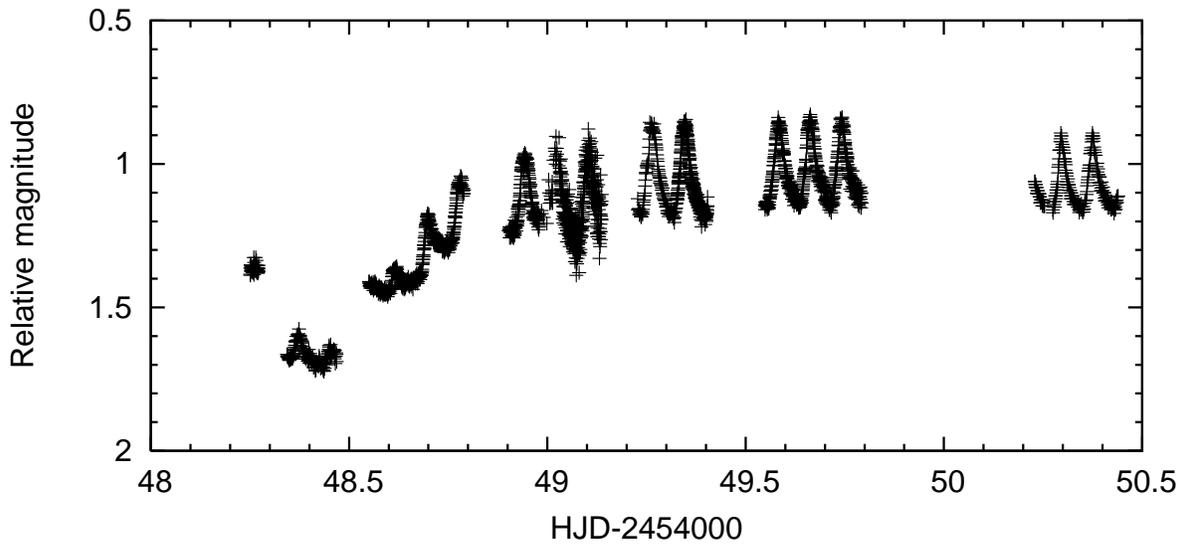}}
\end{center}
\caption{Enlarged light curves around the supermaximum. For the purpose
 of description, data obtained from HJD 2454048.8877 to HJD 2454049.1369
 were binned by 8 frames. Although the data were sparse before HJD
 2454048.5, the light curves provided evidence for a precursor around
 the phase, where a hint of modulations was visible. As can be seen in
 this figure, the supermaximum was around HJD 2454049.7.}
\label{elc}
\end{figure*}

\subsection{light curves}

Figure \ref{lc} illustrates the overall light curves of EG Aqr during
the 2006 outburst. By virtue of the prompt reports of the outburst, the
data covered the early stages before the outburst maximum. Enlarged
light curves around the maximum are displayed in figure
\ref{elc}. Judging from the trend of the light curve, we may have caught
the end of the precursor and rising stage of the main outburst on HJD
2454048. At the end of the precursor, some superhumps were visible. The
magnitude reached the maximum on HJD 2454049, after which the variable
declined at the rate of 0.14(1) mag d$^{-1}$. Based on its quiescent and
maximum magnitude, the superoutburst amplitude exceeded ${\sim}$5.8
mag. From this, we can categorize EG Aqr as large-amplitude dwarf novae
(TOADs, \cite{how95TOAD}). The duration of the plateau stage was 8-11
days. No rebrightenings were observed during our run.

\subsection{superhump}

To clarify EG Aqr as a new member of SU UMa-type dwarf novae, we
examined the light curves obtained at each site. Except for some data
obtained under bad conditions, we can unambiguously detect superhumps
having a rapid rise and slow decline of the profiles. This is the
first time that EG Aqr showed superhumps.

\begin{figure}
\begin{center}
\resizebox{80mm}{!}{\includegraphics{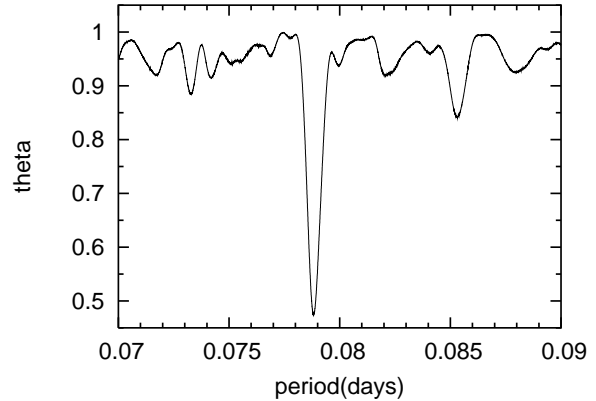}}
\end{center}
\caption{Theta diagram of period analysis for the plateau stage between
 HJD 2454049.55 and HJD 2454056.98. The lowest minimum is definitely
 seen at $P_{\rm sh}$=0.078828(5) days.}
\label{pdmp}
\end{figure}

\begin{figure}
\begin{center}
\resizebox{80mm}{!}{\includegraphics{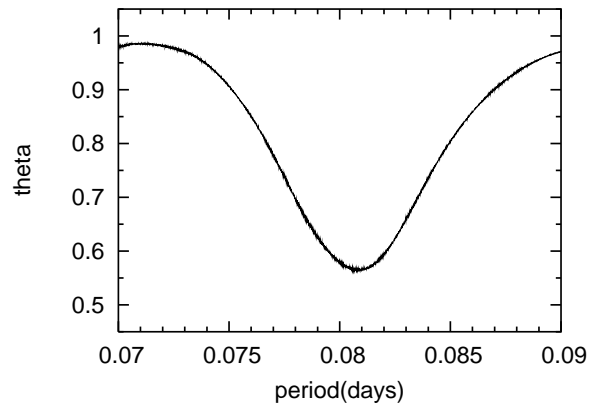}}
\end{center}
\caption{Same as figure \ref{pdmp} but for the data between HJD
 2454048.25 and HJD 2454049.40. The best estimated period during the
 phase yields $P_{\rm sh}$=0.08076(12) days.}
\label{pdme}
\end{figure}

\begin{figure}
\begin{center}
\resizebox{80mm}{!}{\includegraphics{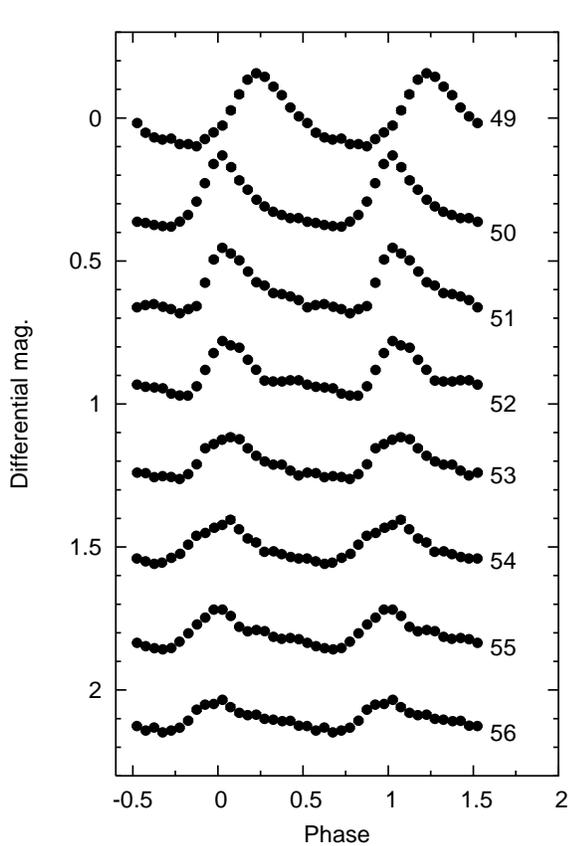}}
\end{center}
\caption{Nightly averaged profiles of superhumps. The numbers in this
 figure denote the days since HJD 2454000. Based on the above estimated
 period, the profile on HJD 2454049 was folded by 0.08076 days, while
 folded by 0.078828 days after HJD 2454050. Although the amplitude of
 the superhumps decreased gradually, the feature of a rapid rise and
 slow decline were visible for all the respective nights.}
\label{mhp}
\end{figure}

After subtracting trends for each light curve, we applied the phase
dispersion minimization method (PDM, \cite{ste78pdm}) to the
residual light curves from HJD 2454049.5 to HJD 2454057, corresponding to
the plateau stage. The resultant theta diagram is displayed in figure
\ref{pdmp}, in which we determined 0.078828(6) days as the best
estimated period of the mean superhump. The error of the period is
calculated using the Lafler-Kinman class of methods, as applied by
\citet{fer89error}. The same method was applied for the data before HJD
2454049.5. from which we determined $P$=0.08076(11) days as being the
best estimated period. Figure \ref{pdme} shows the resultant theta
diagram. These results indicate
the mean superhump period around the maximum was about 0.002 days longer
than that during the main plateau stage. In figure \ref{mhp}, we demonstrate
the averaged profiles of the superhumps folded by 0.08076 days from HJD
2454049 and 0.078828 days for HJD 2454050 through HJD 2454057, where one can
see a characteristic feature of superhumps. Although the amplitude of
the superhumps decreased as the superoutburst proceeded, the shape of
the profiles kept almost constant over the course of the plateau
stage. The apparent absence of eclipses indicates that the inclination
of the system is not too high.

\subsection{superhump period change}

\begin{figure}
\begin{center}
\resizebox{80mm}{!}{\includegraphics{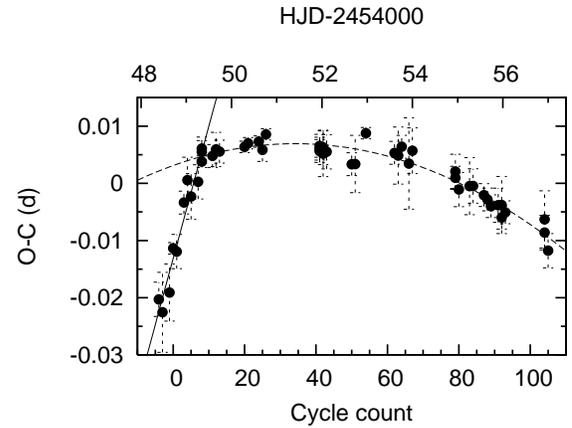}}
\end{center}
\caption{$O-C$ diagram during the superoutburst. The horizontal and
 vertical axes mean $O-C$ and the cycle count, respectively. The maximum
 timings were listed in table 2. A linear regression to the
 observed times gives equation (\ref{eqlinear}). The solid and dashed lines
 indicate the best fit equation for -4$<E<$8 and 11$<E<$110,
 respectively. The former is given by equation (\ref{eqearly}), while
 the latter is given by equation (\ref{eqplateau}). This figure means
 that the period kept almost constant in -4$<E<$8. However, an abrupt
 period change occurred in 8$<E<$11. After $E>$11, the superhump
 period decreased gradually at the rate of $\dot{P}$/$P$ =
 $-$8.2$\times$10$^{-5}$.}
\label{oc}
\end{figure}

We calibrated the maximum timings of superhumps mainly by eye. The
error is an order of 0.001 days. In table 2, we tabulate the
maximum timings of superhumps. A linear regression to the obtained
values in table 2 yielded,

\begin{equation}
HJD(max) = 2454048.7101(18) + 0.078935(29) \times E,
\label{eqlinear}
\end{equation}
where $E$ is the cycle count. Using the equation, we can
draw an $O - C$ diagram described in figure \ref{oc}. It should be noted
that the change of the period occurred for 8${\leq}E{\leq}$11, when we missed
observations unfortunately. Nevertheless, we firstly succeeded in
detecting the superhump period change at the early stage of
superoutburst for long-period SU UMa-type dwarf novae, and confirmed the
period change did occur. For -4$<E<$8, the best fit equation is given by

\begin{equation}
O - C = -1.40(0.09)\times10^{-2} + 2.35(0.18)\times10^{-3} E. 
\label{eqearly}
\end{equation}
On the other hand, the best fit quadratic between $E$=11 and $E$=114 is given by
\begin{eqnarray}
O - C =& 2.37(0.90)\times10^{-3} + 2.52(0.33)\times10^{-4} E \nonumber \\
       & -3.23(0.26)\times 10^{-6} E^{2}. 
\label{eqplateau}
\end{eqnarray}
The equation \ref{eqplateau} indicates $P_{\rm dot}$ = $\dot{P}$/$P$ =
$-$8.2(7)$\times$10$^{-5}$. This is a normal value for SU UMa-type dwarf
novae \citep{2006PASJ...58..383I}.

\subsection{post outburst stage}

\begin{figure}
\begin{center}
\resizebox{80mm}{!}{\includegraphics{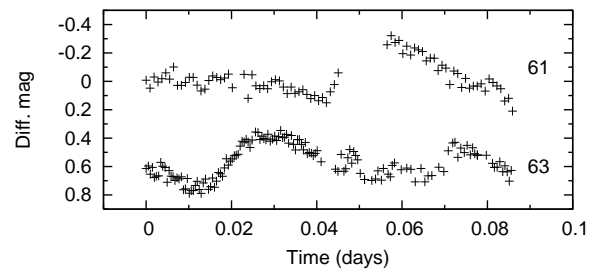}}
\end{center}
\caption{Representative light curves during the post-outburst
 phase. For a visual purpose, these light curves were arbitrarily
 shifted. The numbers in the figure denote the days from 2454000. The
 datapoints are binned by 4 frames on HJD 2454063 (November 24) after
 precluding the bad data. There exist large modulations on both light
 curves.}
\label{postlc}
\end{figure}

Some SU UMa-type dwarf novae exhibit rebrightenings and/or late superhumps
after the main plateau stage. In order to search for these phenomena, we
mainly used 1.5m Kanata telescope. Figure \ref{postlc} shows the
representative light curves after the main plateau stage obtained by the
telescope. One can see modulations with an amplitude larger than 0.3
mag, which is even larger than that of superhumps. On November 24, the
light curve appears to be doubly-peaked indicating the light curve had
remained unchanged. However, due to the lack of observations, it is not
clear whether the observed light curves are late
superhumps. \footnote{We calibrated the maximum timing of the hump on
November 24. However, we cannot draw firm conclusion whether a phase
shift occurred, since the interval between the two successive humps was
too long (see table 2).}

In order to explore whether a rebrightening showed, we examined the
VSNET and AAVSO databases. As can be seen in figure \ref{lc}, there is
no evidence for a rebrightening before HJD 2454065. Although most of
rebrightening are observed within 5 days after the end of the plateau
stage (e.g., \cite{how96alcom}; \cite{ish01rzleo};
\cite{tem06asas0025}), we cannot rule out the possibility that a
rebrightening occurred between HJD 2454066 and HJD 2454070.

\begin{table*}
\caption{Timings of superhump maxima.}
\begin{center}
\begin{tabular}{cccccccc}
\hline\hline
$E^*$ & HJD$^{\dagger}$ & err$^{\ddagger}$ & ID & $E^*$ &
 HJD$^{\dagger}$ & err$^{\ddagger}$ & ID \\
\hline
-4 & 48.3731 & 0.003 & GM & 51 & 52.7400 & 0.005 & KTC\\
-3 & 48.4498 & 0.007 & GM & 54 & 52.9823 & 0.001 & Njh \\
-1 & 48.6112 & 0.005 & KTC & 62 & 53.6106 & 0.002 & KTC\\ 
0 & 48.6979 & 0.001 & KTC & 63 & 53.6891 & 0.005 & KTC\\
1 & 48.7763 & 0.003 & KTC & 64 & 53.7696 & 0.002 & KTC\\ 
3 & 48.9428 & 0.002 & Kyoto & 66 & 53.9246 & 0.008 & Mhh \\ 
4 & 49.0257 & 0.004 & Kis & 67 & 54.0058 & 0.004 & Mhh \\ 
5 & 49.1018 & 0.004 & Mhh & 79 & 54.9487 & 0.002 & Kyoto \\
7 & 49.2623 & 0.003 & GM & 79 & 54.9498 & 0.003 & Njh \\
8 & 49.3448 & 0.001 & AAVSO &  80 & 55.0256 & 0.003 & Njh \\
8 & 49.3465 & 0.002 & GM &  83 & 55.2631 & 0.005 & GM \\
8 & 49.3471 & 0.002 & BM &  84 & 55.3421 & 0.003 & GM \\
11 & 49.5827 & 0.002 & KTC&  87 & 55.5774 & 0.002 & KTC\\
12 & 49.6628 & 0.003 & KTC& 88 & 55.6556 & 0.002 & KTC\\
13 & 49.7414 & 0.002 & KTC& 89 & 55.7334 & 0.002 & KTC\\
20 & 50.2950 & 0.001 & GM & 91 & 55.8915 & 0.003 & Kyoto \\
21 & 50.3746 & 0.001 & GM & 92 & 55.9683 & 0.002 & Kyoto \\
24 & 50.6118 & 0.001 & KTC& 92 & 55.9705 & 0.005 & Njh \\
25 & 50.6893 & 0.002 & KTC& 93 & 56.0482 & 0.002 & Njh \\
26 & 50.7710 & 0.001 & KTC& 101 & 56.6770 & 0.004 & KTC \\
41 & 51.9526 & 0.002 & Njh & 102 & 56.7557 & 0.004 & KTC \\
41 & 51.9530 & 0.002 & Mhh & 104 & 56.9133 & 0.003 & Kyoto \\
41 & 51.9535 & 0.002 & Kyoto & 104 & 56.9156 & 0.005 & Njh \\
42 & 52.0311 & 0.004 & Mhh & 105 & 56.9891 & 0.003 & Kyoto \\
42 & 52.0322 & 0.003 & Njh & 113 & 57.6194 & 0.001 & KTC \\
43 & 52.1104 & 0.003 & Mhh & 114 & 57.6985 & 0.002 & KTC \\                          
50 & 52.6610 & 0.002 & KTC & 194: & 63.9761 & 0.004 & Kanata \\
\hline
\multicolumn{3}{l}{$^*$ Cycle count.} \\
\multicolumn{3}{l}{$^{\dagger}$ HJD-2454000} \\
\multicolumn{3}{l}{$^{\ddagger}$ In the unit of day.} \\
\end{tabular}
\end{center}
\label{hjdmax}
\end{table*}

\section{Discussion}

\subsection{long-term activity}

One of the most unique properties in EG Aqr is the inactivity of the
system. As mentioned before, there is only one recorded outburst since
the discovery of EG Aqr until the current one. No
outbursts were detected by the ASAS-3 over the past 5 years, during
which the variable was below the detection limit of $V$=14.5. We also
explored visual observations reported to the AAVSO and VSNET since
1999. EG Aqr was especially monitored in 1999, 2000, and 2004 with a
mean interval of observations being a few days. These intensive
observation may detect not only superoutburst but also normal outburst
if they really occur. However, no outbursts were reported until 2006.
Although we should await an investigation on archival plates to draw a
firm conclusion as to whether EG Aqr is indeed inactive, the current
available information on EG Aqr may suggest its inactive nature. The
non-detection of a normal outburst in EG Aqr may also support its long
recurrence time, since the frequency of normal outburst is related to
the length of the supercycle \citep{osa95wzsge}.

\subsection{outburst light curve}

As noted above, extensive CCD photometry revealed the SU UMa nature of
EG Aqr. Superhumps were clearly visible during the course of the
superoutburst for the first time in EG Aqr. As for the light curve
itself, we should note on the presence of a precursor and the presence
of superhumps at the end of the precursor. In the previous
section, we have shown the $O-C$ diagram, in which we measured the
maximum timings at the end of the precursor, corresponding to $E$=-4 and
-3. The obtained $O-C$ diagram strongly suggests that the light source
of the variations in -4$<E<$8 comes from the same origin, since a
phase shift or a change of the period were not seen. This means that the
modulations were coherent for -4$<E<$8. As can be seen in figure
\ref{elc} the modulations that we observed around $E$=8 were genuine
superhumps. Therefore, it is likely that superhumps had already grown at
the end of the precursor. If this is the case, EG Aqr is the second
example that superhumps were seen even during a precursor, following the
1993 superoutburst of T Leo \citep{kat97tleo}.

Besides the presence of the superhumps during the precursor, the overall
light curve is typical of SU UMa-type dwarf novae, in terms of
the duration of the plateau phase, as well as the decline rate of the
magnitude. The presence of a precursor also supports its SU UMa nature
of EG Aqr, since no precursor has been reported for WZ Sge-type dwarf
novae \citep{kat04egcnc}.

\subsection{superhump period change}

As is well known, most of SU UMa-type dwarf novae decrease their
superhump period during the course of the plateau stage. This is due to
shrinkage of the disk, or a natural consequence of the depletion of the
mass \citep{osa85SHexcess}. Recent studies, however, have shown that some
systems increase the superhump period during the plateau phase
\citep{nog97alcom}. So far, the exact mechanism that causes the
superhump period change is not known.

Recently, \citet{uem05tvcrv} has suggested that the superhump period change
may be related to the presence or absence of a precursor. The authors
studied a short period SU UMa-type dwarf nova TV Crv for the 2002 and
2004 superoutburst, and discovered that the superhump period increased
during the 2002 superoutburst without a precursor, while hardly changed
during the 2004 superoutburst with a precursor. Theoretically,
\citet{osa03DNoutburst} proposed the refined thermal-tidal instability
model, in
which a superoutburst without a precursor have a larger disk radius
than that with a precursor. When the accretion disk reaches the tidal
truncation radius as a result of the mass accretion onto the white
dwarf, the stored matter at the outer edge may prevent the inner
propagation of the
cooling wave, so that the disk keeps the hot state. On the other hand, if
the accretion disk does not reach the tidal truncation radius, the
outermost region of the disk allows the cooling wave to propagate
inward. As a consequence, the outburst is quenched like a normal
one. If the eccentricity of the accretion disk grows sufficiently,
the tidal dissipation at the rim of the disk leads to the orientation of
the heating wave. This process is observed as a superoutburst with a
precursor.

In the case of EG Aqr, the superoutburst was accompanied by a precursor,
indicating that the radius of the accretion disk did not reach the tidal
truncation radius. This, however, seems to be peculiar when taking into
account the long quiescent time of the object. The long quiescence means
a large amount of mass in the accretion disk, like WZ Sge-type dwarf
novae. However, the duration of the plateau stage, as well as the
presence of a precursor are indicative of the small disk of the object
compared to WZ Sge-type dwarf novae.

During the main plateau stage, the best fitting quadratic formula
yielded $P_{\rm dot}$ = $\dot{P}$/$P$ = $-$8.2$\times$10$^{-5}$, which
is a normal value for SU UMa-type dwarf novae. The value is quite different
from that observed in WZ Sge-type stars, which show
positive period derivatives, or almost constant value of the superhump
periods \citep{ish03hvvir}. Although quiescence of EG Aqr behaves as WZ
Sge-type dwarf novae, the overall feature of the superhump period
changes and the light curve of the superoutburst bear strong resemblance
to those of usual SU UMa-type dwarf novae.

Regarding the early phase of the observations, it again should be noted
that the superhump period unchanged and a sort of period change occurred
near the bright maximum. Although evidence of such period changes has
been provided in other SU UMa-type dwarf novae (e.g., figure 5 of
\citet{uem05tvcrv}), this is the first case that we specify when the
transition occurred. In future, we should investigate whether the period
change is a ubiquitous property for SU UMa-type dwarf novae by observing
other objects, which shed lights on understanding the origin of the
period shift.

\subsection{inactive SU UMa-type dwarf novae with long superhump periods}

So far, as much as 200 objects have been confirmed to be SU UMa-type
dwarf novae, for which we may divide them into two categories
according to their superhump periods and their tendency of the period
derivatives. There exists a rough cutoff around $P_{\rm sh}$=0.063 days
below which the superhump period tends to increase during the superoutburst
\citep{ima05gocom}. On the other hand, for systems above $P_{\rm
sh}$=0.063 days, where the majority of SU UMa-type dwarf novae exists,
the superhump period tends to decrease or keep almost constant. The latter
systems are sometimes introduced as ``textbook'' SU UMa-type dwarf novae
\citep{kat03v877arakktelpucma}.

However, when we focus on the long-term behavior of the respective
system, we notice the diversity of the nature. For example, a prototype
SU UMa-type star VW Hyi shows frequent normal outbursts between two
successive superoutbursts with a mean recurrence time of the superoutburst
being 150 days \citep{osa89suuma}. Such systems are well reproduced by the
thermal-tidal instability model (\cite{osa89suuma};
\cite{osa03DNoutburst}). Some long-period systems, on the other hand, hardly
show not only a normal outburst, but also a superoutburst. These systems
include EF Peg (\cite{how93efpeg}; \cite{kat02efpeg};
\cite{how02llandefpegHST}),
V725 Aql (\cite{1996JAVSO..24...14H}; \cite{uem01v725aql}), QY
Per\footnote{http://www.kusastro.kyoto-u.ac.jp/vsnet/DNe/qyper.html},
and EG Aqr. All of these systems have a long recurrence time of
years, indicating that the mass transfer rate from the secondary is
supposed to be small. This seems to be curious because the secular
evolution of dwarf novae suggests the existence of relatively high mass
secondaries compared to those in short period dwarf
novae unless the binaries are so called ``period bouncers''
(\cite{kin88binaryevolution}; \cite{kol99CVperiodminimum};
\cite{pat05re1255}). It is unlikely that these systems had passed
the period minimum of dwarf novae, since the evolutional timescale for
``period bouncers'' may be as long as ${\sim}$ 10 Gyrs
\citep{pat98evolution}. Optical and infrared spectroscopy show mid
M-type secondaries in EF Peg \citep{how02llandefpegHST}. This indicates
that this system does not reach the period minimum, because the ``period
bouncers'' are believed to contain a brown dwarf secondary
\citep{how97periodminimum}.

At present, we cannot specify the reason why such inactive systems do
exist in long period systems. In fact, these systems are less studied,
especially during quiescence. Actually, we have no information on the
orbital period of the above four systems.\footnote{\citet{how93efpeg}
determined a likely orbital period of EF Peg as 2.05 hours based on the
observed superhump period. It should be noted that this is not a direct
measurement of the orbital period.} In future, quiescent spectroscopy is
imperative in order to determine the orbital period and the spectral
type of the secondary of the system. This would shed some light
on the evolutional status of the above mentioned systems.

\section{Conclusion}

We established the SU UMa nature of EG Aqr for the first time by the
detection of the superhumps with a mean period of 0.078828(6)
days. The observed superoutburst showed a precursor, during which a hint
of superhumps was seen at the last stage of the precursor. Extensive
observations enabled us to examine detailed changes of the superhump
period. It turned out that the superhump period kept almost constant
near the bright maximum, after which the superhump period decreased
normally at the rate of $P_{\rm dot}$ = $\dot{P}$/$P$ =
$-$8.2$\times$10$^{-5}$ all over the plateau stage. Although the origin
is unknown, we detected a change of the period near the bright
maximum. Concerning the observed period change and the presence of the
precursor, the maximum radius of the accretion disk was not large during
the superoutburst. The obtained light curves were typical of those of SU
UMa-type dwarf novae. This is definitely peculiar when taking into
account that EG Aqr showed only one recorded eruption until the 2006
superoutburst. Despite the inactivity of the variable, we conclude that
EG Aqr is a new member of SU UMa-type dwarf novae. In future, quiescent
studies both from photometry and spectroscopy are imperative in order to
further understand the enigmatic object.

\vskip 3mm

We would like to thank Dr. Steve B. Howell for helpful comments on the
paper. We acknowledge with thanks the variable star observations
from the AAVSO and VSNET International Database contributed by observers
worldwide and used in this research. We would also express our gratitude
to G. Pojmanski for providing invaluable data of ASAS-3
observations. This work is supported by a Grant-in-Aid for the 21st
Century COE ``Center for Diversity and Universality in Physics'' from
the Ministry of Education, Culture, Sports, Science and Technology
(MEXT). GM acknowledges the support of the Planetary Society and
Software Bisque. This work is partly supported by a grant-in aid from
the Ministry of Education, Culture, Sports, Science and Technology
(No. 14079206, 16340057, 17684004, 17340054, 17740105, 18740153,
18840032). Part of this work is supported by a Research Fellowship of
the Japan Society for the Promotion of Science for Young Scientists (RI,
AI, KK, ON).

\end{document}